\documentclass[aps,showpacs,twocolumn]{revtex4}
\usepackage{bm}
\usepackage{graphicx}
\usepackage{epsfig}
\begin{document}

\title{Intervalley Scattering and Weak Localization in Si-based
Two-Dimensional Structures}

\author{A.\,Yu.\,Kuntsevich$^{a}$, N.\,N.\,Klimov$^{b}$, S.\,A.\,Tarasenko$^c$,
N.\,S.\,Averkiev$^c$, V.\,M.\,Pudalov$^{a}$, H.\,Kojima$^{b}$, and
M.\,E.\,Gershenson$^{b}$}

\address{$^{a}$ P.\,N.\,Lebedev Physics
Institute, 119991 Moscow, Russia }
\address{$^b$ Department of Physics and
Astronomy, Rutgers University, New Jersey 08854, USA}
\address{$^c$ A.\,F.\,Ioffe Physico-Technical Institute, 194021 St.Petersburg, Russia}

\begin{abstract}
We have measured the weak localization magnetoresistance in
(001)-oriented  Si MOS structures with a wide range of mobilities.
For the quantitative analysis of the data, we have extended the
theory of weak-localization corrections in the ballistic regime to
the system with two equivalent valleys in electron spectrum. This
theory describes the observed magnetoresistance and allows the
extraction of the phase breaking time $\tau_{\varphi}$ and the
intervalley scattering time $\tau_{\rm v}$. The temperature
dependences $\tau_{\varphi}(T)$ for all studied structures are in
good agreement with the theory of electron-electron interaction
effects in two-dimensional systems. The intervalley scattering is
elastic and rather strong: $\tau_{\rm v}$ is typically only an
order of magnitude greater than the transport time, $\tau$. It is
found that the intervalley scattering rate is
temperature-independent and the ratio $\tau_{\rm v}/\tau$
decreases with increasing the electron density. These observations
suggest that the roughness of the Si-SiO$_2$ interface plays the
major role in intervalley scattering.
\end{abstract}


\date{\today}

\maketitle

\section{Introduction}
Two-dimensional semiconductor structures with a degenerate ground
state attract a great deal of attention because of the richness of
low-temperature transport and thermodynamic phenomena that are
often absent in systems with a simple band structure. The energy
spectrum of a two-dimensional electron system in the
metal-oxide-semiconductor (MOS) structures grown on the
(001)-oriented Si surface consists of six subbands (valleys). At
low temperatures and low electron densities, only two of them are
occupied (the other four valleys are considerably higher in
energy) \cite{ando}. The low-energy valleys are almost equivalent:
the valley splitting $\Delta_{\rm V}$ caused by an asymmetry of
the confining potential is typically negligible in comparison with
the Fermi energy \cite{ando}. When the intervalley scattering is
weak, the valley degeneracy strongly affects both
electron-electron interaction and weak localization (WL) effects
in the conductivity. In particular, the interaction effects in Si
MOS structures are strongly amplified by the valley degeneracy
\cite{finkel'steinPRL}; this accounts for the anomalous
``metallic'' temperature dependence of the resistivity in
high-mobility Si MOSFETs at intermediate temperatures
\cite{ZNA,pudalovmetallic}. Accordingly, the intervalley
scattering plays an important role in the low-temperature
phenomena in Si MOS structures: it determines the low-temperature
cut-off of the metallic-like transport and could also modify the
2D metal-insulator transition observed in these structures at low
electron densities \cite{kravchenkoMIT,finkel'stein}. However, to
the best of our knowledge, there were no systematic studies of the
intervalley scattering in Si MOS structures.

The measurements of weak localization corrections to the
conductivity of two-valley systems allow one to study intervalley
scattering. The effect of intervalley scattering on WL depends on
the relationship between the intervalley scattering time,
$\tau_{\rm v}$, and the time of dephasing of the electron wave
function, $\tau_{\varphi}$.  For weak intervalley scattering
($\tau_{\rm v}>>\tau_{\varphi}$), two valleys are independent at
the time scale $\tau_{\varphi}$ relevant to the diffusion regime
of WL. It is therefore expected  that the magnitude of the WL
correction is doubled in comparison with its value in a system
with strong intervalley scattering ($\tau_{\rm
v}<<\tau_{\varphi}$); the latter correction is the same as in a
single-valley system because the valleys are completely mixed at
the $\tau_\varphi$ time scale.

In numerous measurements of the WL magnetoresistance in Si MOS
structures \cite{wheeler,kawaji,pudalovwl,kravchenkowl},  the
experimental data were fitted using the Hikami-Larkin-Nagaoka
(HLN) theory \cite{hikami}. Interestingly, the factor-of-two
enhancement of the WL correction was never observed, indicating
that intervalley scattering is rather strong. In order to extract
the intervalley scattering time from the WL magnetoresistance, the
measurements should be extended towards higher magnetic fields.
However, the HLN theory, which is used for fitting the WL
magnetoresistance, was developed within the diffusive
approximation (i.e small magnetic fields, see below). Therefore,
for an adequate description of the effect of intervalley
scattering on WL in Si MOS structures, a theory applicable over a
wider range of magnetic fields should be developed.

In this paper, we extend the ballistic (i.e. applicable to
arbitrary classically-weak magnetic fields) theory of WL
corrections to the case of two degenerate valleys. This enables
the detailed quantitative analysis of the WL magnetoresistance
measured for several Si MOS structures with the electron mobility
varying over an order of magnitude.  Rather small extracted values
of $\tau_{\rm v}$ ($\sim 10 \tau$) indicate that the intervalley
scattering in Si MOS structures is strong. We have found that (i)
intervalley scattering is temperature-independent, i.e. elastic,
(ii) the scattering rate depends monotonically on the electron
density, and (iii) there is no simple correlation between the
intervalley scattering and the mobility for different samples. The
phase relaxation time in all studied structures is well described
by the theory of electron-electron interaction
effects~\cite{znatauphi}.

The paper is organized as follows. In section \ref{expsection} we
describe the samples, measurement techniques and provide examples
of experimental data. In section \ref{firstprocessing}, to compare
our data with previously reported results, we fit the WL
magnetoresistance with the HLN theory. The results of this
analysis indicate that the WL correction is approximately a factor
of 2 smaller than one could expect for a system with two
independent valleys. The theory of the WL magnetoconductance for
systems with degenerate multi-valley spectrum is presented in
section \ref{theory} and the data are analyzed using this theory
in section \ref{ballisticprocessing}. In section \ref{results} we
discuss the results of the data analysis, in particular, the
intervalley and phase breaking times.

\section{Experiment}
\label{expsection} Below we present data for three representative
Si MOSFET samples: (i) high-mobility ($\mu\approx2$~m$^2$/Vs at
0.1\,K) sample Si6-14 \cite{pudalovmetallic} which demonstrates a
strongly pronounced ``metallic-like'' dependence $\sigma(T)$ and a
metal-insulator transition with decreasing electron density $n$,
(ii) sample Si39 with an intermediate mobility
$\mu\approx0.45$~m$^2/$Vs \cite{si39example}, and (iii)
low-mobility sample Si40 with $\mu\approx0.18~$m$^2$/Vs at
$T<4.2$\,K. The transport times for these samples within the
studied range of $n$ were $\tau\approx2$~ps, $0.6$~ps, and
$0.2$~ps, respectively. The resistance was measured using the
standard four-terminal AC technique by a resistance bridge LR700
(for sample Si6-14) and by SR830 lock-in amplifier in combination
with a differential preamplifier (for samples Si39 and Si40). A
sufficiently small measuring current (1-3~nA for Si6-14 and
50--100~nA for Si39 and Si40) was chosen to avoid overheating of
electrons. The electron density, controlled by the gate voltage,
was found from the period of the Shubnikov-de Haas oscillations
and the dependence of the Hall resistance on magnetic field. Both
results were consistent with each other within 2\% accuracy in the
studied range of densities; this uncertainty of $n$ is
insignificant for the further analysis. To ensure that only the
lowest size quantization subband is filled \cite{ando}, we
performed measurements at $n<4\times10^{12}$cm$^{-2}$.
\begin{figure}
 \centerline{\psfig{figure=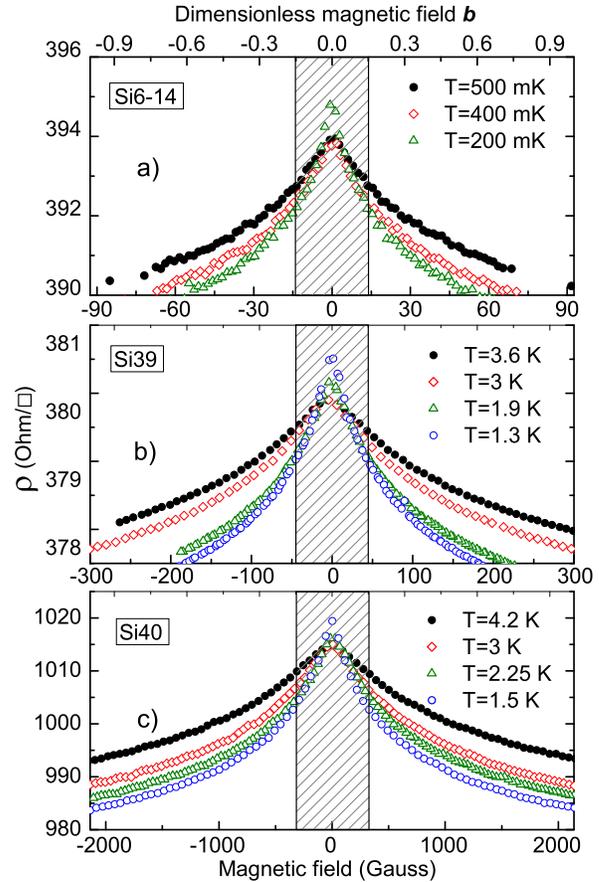,width=260pt}}
\caption{Examples of the magnetoresistance $\rho(B)$ data for
Si6-14, $n=1\times 10^{12}$cm$^{-2}$(a), Si39, $n=3.6\times
10^{12}$cm$^{-2}$ (b), and Si40, $n=3.5\times
10^{12}$cm$^{-2}$(c). The data within the hatched regions have
been used to extract the $\tau_\varphi$ value. Upper axes show the
magnetic field in units of $B_{tr}={\Phi_0}/{2\pi l^2}$, lower
axes show the field in Gauss. The data points were chopped for
clarity.} \label{fig1}
\end{figure}
\vspace{0.15in}

The WL magnetoresistance was measured at $T=0.05-0.6$\,K for
high-mobility sample Si6-14, and at $T=1.3-4.2$\,K for samples
Si39 and Si40. At these temperatures,  the phase breaking time
exceeds the transport time by one to two orders of magnitude. The
magnetic field aligned perpendicular to the plane of Si MOS
structures was varied from -1 to +1\,kG (Si6-14) and from -3 to
+3kG (Si39 and Si40). When sample Si6-14 was measured at $T<1$\,K,
an additional in-plane field $\sim$200\,G was applied to quench
the superconductivity in the current/voltage contact pads and the
gate electrode made of thin aluminum film. For reliable extraction
of the phase relaxation time $\tau_\varphi$ from the WL
magnetoresistance, we have chosen a small field step size: 1\,G
for Si6-14 and 3\,G for Si39 and Si40. The examples of
magnetoresistance $\rho(B)$ data for Si6-14, Si39 and Si40 at a
fixed density and various temperatures are shown in Figs.
\ref{fig1}a, \ref{fig1}b and \ref{fig1}c, respectively. Hereafter
throughout the paper we will use magnetoconductance (MC) $\Delta
\sigma\equiv\rho(B)^{-1}-\rho(B=0)^{-1}$.

\section{Fitting the data with the Hikami-Larkin-Nagaoka theory}
\label{firstprocessing}

\begin{figure}
\vspace{-0.15in}
 \centerline{\psfig{figure=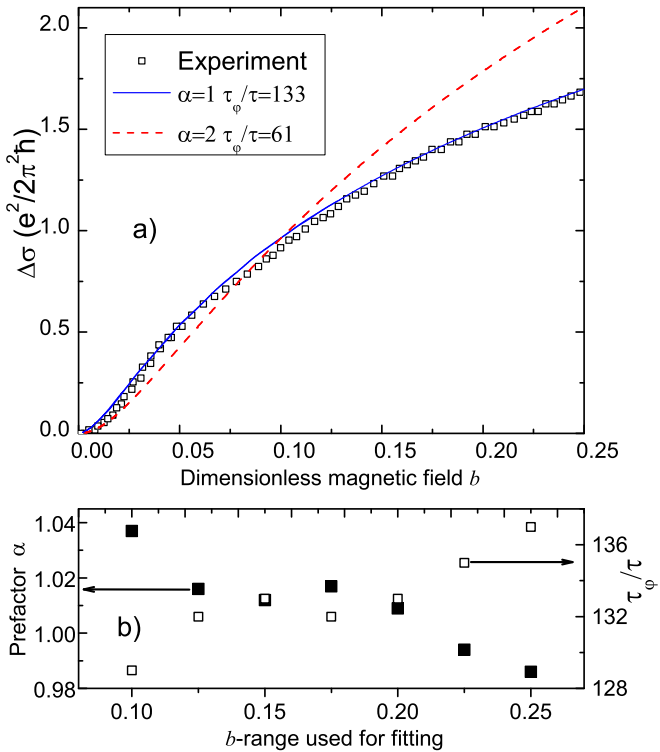,width=280pt}}
 \caption {a) Example of the MC
data (points) for sample Si40, $T=1.45K$ and
$n=$3.34$\times10^{12}$cm$^{-2}$. Solid line is calculated using
Eq.(1) with two fitting parameters $\alpha=1$ and $\tau_\varphi /
\tau=133$. The dashed curve is an attempt to fit the same data
over the same range of $b$ with a fixed prefactor $\alpha=2$ and
$\tau_\varphi / \tau$ =61. b) Dependences of the fitting
parameters $\alpha$ (solid squares) and $\tau_\varphi / \tau$
(open squares) on the magnetic field range $0$ -- $b$ which was
used for fitting. } \label{fig2}
\end{figure}

It is a common practice to extract the phase breaking time from
the WL magnetoconductance using the HLN
theory\cite{wheeler,kawaji,pudalovwl,kravchenkowl,minkovgornyi}:
\begin{equation}
\label{HikamiFormula} \Delta \sigma_{\rm
HLN}\left(b,\frac{\tau_{\varphi}}{\tau}\right)=\frac{\alpha
e^2}{2\pi^2\hbar}\left[
\psi\left(\frac{1}{2}+\frac{\tau}{b\tau_{\varphi}}\right)+\ln{\frac{b\tau_{\varphi}}{\tau}}\right].
\end{equation}
Here $\psi$ is the digamma-function, $e$ is the electron charge,
$\hbar$ is the Planck constant, $b=B/B_{tr}$ is the dimensionless
magnetic field, $B_{tr}={\Phi_0}/{2\pi l^2}$, $\Phi_0={\pi \hbar
}/e$, and $l$ is the transport mean free path \cite{ldeterm}. The
prefactor $\alpha$ and the dimensionless ratio
${\tau_\varphi}/{\tau}$ are treated as fitting parameters. Note
that with an increase of the magnetic field, the crossover from
the diffusive regime ($b<<1$) to the ballistic regime ($b\sim 1$)
is expected in the WL corrections. Equation~(\ref{HikamiFormula})
with prefactor $\alpha=1$ is the exact result for a single-valley
system in the diffusive regime, i.e. at $\tau_{\varphi}\gg\tau$
and for sufficiently small magnetic fields $b\ll 1$ \cite{hikami}.
On the other hand, the experimental data are often obtained beyond
these limits and, therefore, should be described by more general
ballistic theories \cite{wittman,dmitriev}.

In Ref. \cite{germanenko}, the HLN theory was numerically compared
with the ballistic theory for a single-valley system for various
values of ${\tau_\varphi}/{\tau}$ and magnetic fields $b$. It was
found that both approaches agree with each other within a limited
range of fields $b<0.15$ provided that ${\tau_\varphi}/{\tau}>30$
and the conductivity is much greater than $e^2/2\pi ^2 \hbar$.
Thus, within these limits, Eq.~(\ref{HikamiFormula}) can be used
for extraction of $\tau_{\varphi}$ from the WL magnetoresistance
in a single-valley system, {and} the adjustable parameter $\alpha$
is aproximately equal to 1. For a system with two valleys and weak
intervalley scattering, the prefactor $\alpha$ is expected to be
two times larger, because each valley contributes the term $\Delta
\sigma_{\rm HLN}$ with $\alpha\approx1$ to $\Delta \sigma$.

Figure \ref{fig2}\,a shows a fit of our typical WL MC curve with
Eq.\,(\ref{HikamiFormula}). The fitting performed over the
magnetic field range $b=0$ -- $0.2$ gives
${\tau_\varphi}/{\tau}=133$ and, contrary to the expectation for a
two-valley system without intervalley scattering, $\alpha=1$.
Changing the magnetic field range, where the data are fitted,
causes only minor variations of these parameters (see
Fig.~\ref{fig2}\,b). An attempt to analyze the MC curve using
Eq.(\ref{HikamiFormula}) with a fixed prefactor $\alpha=2$ results
in a much worse fit (dashed line in Fig.~\ref{fig2}\,a).

Several reasons for the reduction of $\alpha$ in a single valley
system have been considered in Ref. \cite{minkovgornyi}, including
(i) Maki-Thompson correction, (ii) Density-of-states correction,
(iii) higher order corrections in $(k_{\rm F}l)^{-1}$,  where
$k_F$ is the Fermi wave vector, and (iv) low $\tau_\varphi/\tau$
ratio. The corrections (i) and (ii) were shown to be small
\cite{minkovgornyi}. To ensure that the higher order corrections
are also small, we have studied the WL MC only for large
conductances $\sim100 \times e^2/2\pi^2 \hbar$. When the ratio
$\tau_\varphi/\tau$  decreases, the HLN theory becomes inadequate,
and the fitting procedure results in an artificially reduced
prefactor. Correspondingly, we performed measurements at such low
temperatures that the inequality $\tau_\varphi/\tau>30$ was
satisfied. We conclude therefore that the aforementioned reasons
cannot account for a low value of the prefactor $\alpha \approx 1$
in the studied multi-valley structures. It should be noted that
$\alpha\approx 1$ in Si MOS structures was obtained in numerous
previous experiments \cite{wheeler,kawaji,pudalovwl,kravchenkowl}.
We show below that the prefactor reduction can be well described
by the theory which explicitly takes the intervalley scattering
into account.

\section{Theory}
\label{theory} A consistent theory of weak localization is
developed in the framework of the diagram technique. The
weak-localization corrections to the conductivity arise in the
first order in the parameter $(k_F l)^{-1}$, where the mean free
path $l$ is governed by the scattering time $\tau$, which is
controlled by both \textit{intra}-valley ($\tau_{\rm v}$) and
\textit{inter}-valley ($\tau_{i}$) scattering processes:
\begin{equation}
1/\tau = 1/\tau_{\rm v} + 1/\tau_{i} \:.
\end{equation}

The weak-localization correction to the conductivity in the
magnetic field has the form

\begin{equation}\label{sigma_gen}
\Delta\sigma(B) = \Delta\sigma^{(a)} + \Delta\sigma^{(b)} \:,
\end{equation}

where the terms $\Delta\sigma^{(a)}$ and $\Delta\sigma^{(b)}$
correspond to the standard diagrams, which have been considered in
detail in Refs.~\cite{Gasparian85,dmitriev,averkiev}. We neglect
both valley and spin-orbit splitting in Si MOS structures.
\cite{deltav} Then, for the short-range scattering potential
\cite{shortrange}, one obtains

\begin{equation}
\Delta\sigma^{(a)} = \frac{\hbar}{\pi} \sum_{\alpha\beta} \int
J_{x}^2(\bm{\rho},\bm{\rho}')
{\mathcal{C}}_{\beta\alpha}^{(3)\alpha\beta}(\bm{\rho}',\bm{\rho})
\, d\bm{\rho} \,d\bm{\rho}' \:,
\end{equation}

\[
\Delta\sigma^{(b)} = \frac{\hbar}{\pi}
\sum_{\alpha\beta\gamma\delta} \int
J_{x}(\bm{\rho},\bm{\rho}')J_{x}(\bm{\rho}'',\bm{\rho})
W_{\gamma\delta}^{\alpha\beta}
{\mathcal{C}}_{\beta\gamma}^{(2)\delta\alpha}(\bm{\rho}',\bm{\rho}'')
\]

\vspace{-0.5cm}

\[
\times [G^R(\bm{\rho},\bm{\rho}')G^R(\bm{\rho}'',\bm{\rho}) +
G^A(\bm{\rho},\bm{\rho}')G^A(\bm{\rho}'',\bm{\rho})] d\bm{\rho}
d\bm{\rho}' d\bm{\rho}'' \,.
\]

Here $\alpha, \beta ,\gamma, \delta = 1,2$ are the valley indices;
$J_{x}(\bm{\rho},\bm{\rho}')$ is the $x$-component of the current
vertex, which is defined as

\begin{equation}
\bm{J}(\bm{\rho},\bm{\rho}') = i e \frac{k_F \tau}{m^*}
\frac{\bm{\rho}-\bm{\rho}'}{|\bm{\rho}-\bm{\rho}'|}
[G^A(\bm{\rho},\bm{\rho}')+G^R(\bm{\rho},\bm{\rho}')] \:,
\end{equation}

$G^{A}(\bm{\rho},\bm{\rho}')$ and $G^{R}(\bm{\rho},\bm{\rho}')$
are the advanced and retarded Green functions,

\begin{equation}
G^{R(A)}(\bm{\rho},\bm{\rho}') = \sum_{s,k_y}
\frac{\psi_{s,k_y}(\bm{\rho})\psi_{s,k_y}^*(\bm{\rho}')}{E_F-E_s \pm
i \hbar /(2\tau) \pm i \hbar /(2\tau_{\varphi})} \:, \label{eq6}
\end{equation}

$\psi_{s,k_y}(\bm{\rho})$ is the wave function of an electron
subject to an external magnetic field $\bm{B}$, which is
perpendicular to the 2D plane, $s$  and $k_y$ are the quantum
numbers ($s$ is the Landau level number and $k_y$ is the in-plane
wave vector), $E_s=\hbar\omega_c(s+1/2)$ is the energy of the
s$^{\mathrm{th}}$ Landau level, $\omega_c=eB/m^*$ is the cyclotron
frequency, $m^*$ is the effective electron mass;
${\mathcal{C}}_{\gamma\delta}^{(2)\alpha\beta}(\bm{\rho},\bm{\rho}')$
and
${\mathcal{C}}_{\gamma\delta}^{(3)\alpha\beta}(\bm{\rho},\bm{\rho}')$
are the Cooperons which depend on four valley indices. The
parameters $W_{\gamma\delta}^{\alpha\beta}$ are determined by
intervalley and intravalley correlators of the scattering potential
and are defined by

\begin{equation}
\langle V_{\alpha\bm{k}_1,\beta\bm{k}_2}
V_{\gamma\bm{k}_3,\delta\bm{k}_4} \rangle N_{\mathrm{imp}} =
W_{\gamma\delta}^{\alpha\beta}\,
\delta_{\bm{k}_1+\bm{k}_3,\bm{k}_2+\bm{k}_4} \:,
\end{equation}

where $V_{\alpha\bm{k}_1,\beta\bm{k}_2}$ is the matrix element of
scattering between electron states $(\beta,\bm{k}_2)$ and
$(\alpha,\bm{k}_1)$ in zero magnetic field, $\bm{k}_j$
($j=1\ldots4$) are wave vectors in the 2D plane,
$N_{\mathrm{imp}}$ is the two-dimensional density of impurities,
and the angle brackets stand for the averaging over impurity
spatial distribution.

The Cooperons
${\mathcal{C}}_{\gamma\delta}^{(2)\alpha\beta}(\bm{\rho},\bm{\rho}')$
and
${\mathcal{C}}_{\gamma\delta}^{(3)\alpha\beta}(\bm{\rho},\bm{\rho}')$
represent the sums of internal parts of the fan diagrams starting
with two and three lines,
respectively,~\cite{Gasparian85,dmitriev,averkiev}. They can be
found from the following equations:

\begin{equation}
{\mathcal{C}}_{\gamma\delta}^{(2)\alpha\beta}(\bm{\rho},\bm{\rho}')
= \sum_{\nu\mu} W_{\gamma\mu}^{\alpha\nu} \,
W_{\mu\delta}^{\nu\beta} \, P(\bm{\rho},\bm{\rho}')
\end{equation}

\vspace{-0.5cm}

\[
+ \sum_{\nu\mu} W_{\gamma\mu}^{\alpha\nu} \int
P(\bm{\rho},\bm{\rho}'') \,
{\mathcal{C}}_{\mu\delta}^{(2)\nu\beta}(\bm{\rho}'',\bm{\rho}')
\,d\bm{\rho}'' \:,
\]

\[
{\mathcal{C}}_{\gamma\delta}^{(3)\alpha\beta}(\bm{\rho},\bm{\rho}')
=
{\mathcal{C}}_{\gamma\delta}^{(2)\alpha\beta}(\bm{\rho},\bm{\rho}')
- \sum_{\nu\mu} W_{\gamma\mu}^{\alpha\nu} \,
W_{\mu\delta}^{\nu\beta} \, P(\bm{\rho},\bm{\rho}') \:,
\]

where
$P(\bm{\rho},\bm{\rho}')=G^{A}(\bm{\rho},\bm{\rho}')G^{R}(\bm{\rho},\bm{\rho}')$.

To calculate the weak-localization correction to the conductivity
in multi-valley structures, we assume that the impurity potential
is the same for particles in different valleys and, thus, the
electron scattering in the valleys is strongly correlated. In
particular, in the (001)-oriented Si-based structures the nonzero
correlators are

\begin{equation}
W_{11}^{11}=W_{22}^{22}=W_{11}^{22}=W_{22}^{11} \:,\;\;
W_{12}^{21}=W_{21}^{12} \:.
\end{equation}

Other correlators vanish due to averaging over the impurity
positions because the lowest conduction-band valleys in silicon
are located in the $\Delta$-points of the Brillouin zone and,
therefore, the Bloch functions contain oscillatory factors. We
note that the intravalley and intervalley scattering times are
determined by these correlators: $1/\tau_{i} = m^* W_{11}^{11}
/\hbar^3$, $1/\tau_{\rm v} = m^* W_{12}^{21} /\hbar^3$ .

Using the standard procedure (see
Refs.~\cite{Kawabata84,Gasparian85}), one can expand
$P(\bm{\rho},\bm{\rho}')$ and the Cooperons in the series of
eigenfunctions of a particle with the double charge in a magnetic
field and derive equations for the weak-localization corrections
$\Delta\sigma^{(a)}$ and $\Delta\sigma^{(b)}$. Calculations show
that the corrections have the form

\begin{equation}\label{sigma_a}
\Delta\sigma^{(a)} = - \frac{e^2 \,b}{2 \pi^2\hbar}
\sum\limits_{N=0}^{\infty} {\mathcal{C}}_N P^2_N
 \:,
\end{equation}

\begin{equation}\label{sigma_b}
\Delta\sigma^{(b)} = \frac{e^2 \,b}{2 \pi^2\hbar}
\sum\limits_{N=0}^{\infty} ({\mathcal{C}}_N + {\mathcal{C}}_{N+1})
\, Q^2_N/2 \:.
\end{equation}

\begin{equation}
{\mathcal{C}}_N = \frac{2(1-\tau/\tau_{\rm v})^3\,
P_N}{1-(1-\tau/\tau_{\rm v})P_N} + \frac{P_N}{1-P_N} -
\frac{(1-2\tau/\tau_{\rm v})^3 P_N}{1-(1-2\tau/\tau_{\rm v})P_N}
\:,
\end{equation}

where $P_N$ and $Q_N$ are coefficients which are given by

\begin{equation}
P_N = \sqrt{\frac{2}{b}} \int\limits_{0}^{\infty} \exp{\left[ - x
\sqrt{\frac{2}{b}} \left(1+\frac{\tau}{\tau_{\varphi}}\right) -
\frac{x^2}{2} \right]} \mathrm{L}_N(x^2) \, dx \:,
\end{equation}

\[
Q_N = \sqrt{\frac{2}{b}} \int\limits_{0}^{\infty} \exp{\left[ -
x\sqrt{\frac{2}{b}} \left(1+\frac{\tau}{\tau_{\varphi}}\right) -
\frac{x^2}{2} \right]} \frac{\mathrm{L}_N^1(x^2) x}{\sqrt{N+1}}
\,dx \:,
\]

and $\mathrm{L}_N$ and $\mathrm{L}_N^1$ are the Laguerre
polynomials.

Equations~(\ref{sigma_gen}), (\ref{sigma_a}), and (\ref{sigma_b})
describe the WL magnetoconductance over the whole range of
classically weak magnetic fields $\omega_c\tau \equiv \mu B <1$.
In the limit of vanishing intervalley scattering ($1/\tau_{\rm
v}=0$), Eqs.~(\ref{sigma_a}) and (\ref{sigma_b}) are reduced to
the conventional expressions for the WL corrections to the
conductivity of a single-valley system~\cite{Gasparian85} and, in
particular, to the HLN formula~\cite{hikami} in the diffusion
regime. The only difference is a prefactor of 2, which accounts
for the valley degeneracy.

\begin{figure}
\vspace{-0.15in}
 \centerline{\psfig{figure=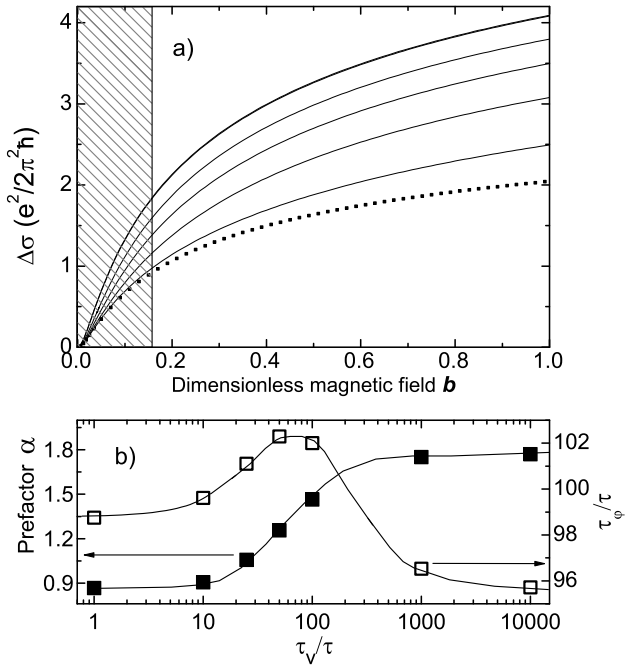,width=260pt}}
\caption {a) WL magnetoconductance calculated for a two-valley
system using Eqs.~\ref{sigma_gen},\ref{sigma_a},\ref{sigma_b}
(solid lines) and for a single-valley system
\protect{\cite{dmitriev}} (dotted line). For solid curves from
bottom to top, $\tau_{\rm v}/\tau$ = 10, 25, 50, 100, 1000, 10000,
respectively. Two upper curves are indistinguishable by eye. For
all the curves $\tau_\varphi/\tau=100$. The hatched region was
used for fitting with Eq.~(\protect\ref{HikamiFormula}). b)
Dependences of the fitting parameters, $\alpha$ and
$\tau_\varphi/\tau$, on $\tau_{\rm v}/\tau$. The data points at
$\tau_{\rm v}/\tau=1$ correspond to a single-valley system.}
\label{fig3}
\end{figure}
\vspace{0.15in}

To illustrate the effect of intervalley scattering on the
magnetoconductance, we calculated the $\Delta\sigma (b)$ dependence
using Eqs.~(\ref{sigma_gen}), (\ref{sigma_a}), and (\ref{sigma_b})
for a fixed $\tau_\varphi/\tau =100$ and various values of
$\tau_{\rm v}/\tau$. The results are shown in Fig.~\ref{fig3}\,a by
solid lines. For comparison, we also calculated the MC using a
similar theory \cite{dmitriev} developed for a single-valley system
(dotted line). We then fitted these dependences over the range
$b<0.15$ using the HLN theory [Eq.~(1)] with two fitting parameters,
the prefactor $\alpha$ and $\tau_\varphi/\tau$. In other words, we
fitted the theoretical curve the same way as the experimental data
have been fitted above in Sec \ref{firstprocessing}.

Figure~\ref{fig3}\,b shows the resultant fitting parameters; for
completeness, we also depicted $\alpha$ and $\tau_\varphi/\tau$
for a single-valley system at $\tau_{\rm v}/\tau=1$. The main
results of the fit are as follows: (i) the extracted phase
breaking time $\tau_\varphi$ coincides with its true value within
a few percent (this uncertainty is insignificant for further
analysis), and (ii) the observed prefactor increases from $\approx
1$ to $\approx 2$ as $\tau_{\rm v}$ increases and becomes greater
than $\tau_\varphi$. {\it Therefore, the approximate equality
$\alpha\approx1$ is simply a consequence of a large ratio
$\tau_\varphi/\tau_{\rm v}\gg 1$}.

\section{Fitting the data with the ballistic theory}
\label{ballisticprocessing} It is intuitively clear and will be
discussed in more detail below, that the MC in low fields $b\ll1$
is predominantly determined by $\tau_{\varphi}$. In principle, one
could use the ``ballistic'' theory for fitting the MC data in the
whole range of magnetic fields and thus for determining both
$\tau_{\varphi}$ and $\tau_{\rm v}$ from a single fit. However the
series Eqs.~(\ref{sigma_a}) and (\ref{sigma_b}) converge very
slowly in small $b$ region. Therefore, to determine
$\tau_{\varphi}$, it is more practical to use in low fields
Eq.~(\ref{HikamiFormula}), that was shown (see Fig.~\ref{fig3}) to
provide the correct $\tau_{\varphi}$ value.

\begin{figure}
\vspace{-0.15in}
 \centerline{\psfig{figure=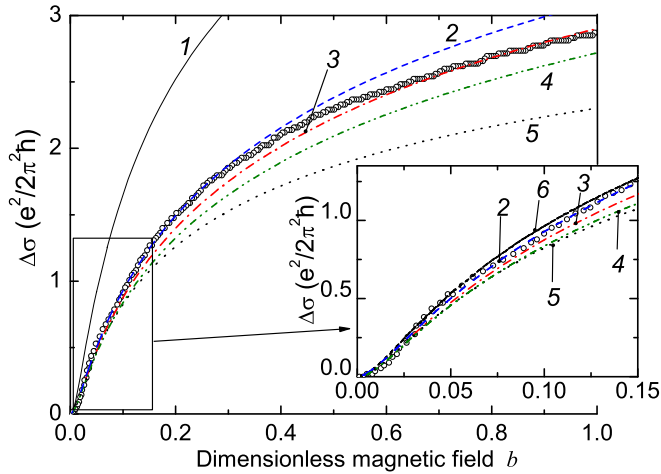,width=280pt}}
\caption {Comparison between the WL magnetoresistance for sample
Si40, $n=33.4\times 10^{11}$cm$^{-2}$, $T=1.45$\,K and the
ballistic theory. Different MC curves are calculated using
Eqs.~\protect{(\ref{sigma_gen}),(\ref{sigma_a}),(\ref{sigma_b})}:
{\it 1} -- for two unmixed valleys ($\tau_{\rm v}=\infty$); {\it
2}~-- $\tau_{\rm v}/\tau=15$; {\it 3}~-- $\tau_{\rm v}/\tau=12$;
{\it 4}~-- $\tau_{\rm v}/\tau=9$; {\it 5}~-- MC for a
single-valley system (equal to curve~{\it 1} divided by 2). Inset
blows up the data in the range $b<0.15$. Curve {\it 6} is the HLN
theory (Eq.~(\protect\ref{HikamiFormula})) with a prefactor
$\alpha=1$ (see Fig.~\protect{\ref{fig2}}\,a). $\tau_\varphi/\tau
=133$ for all calculated curves.} \label{fig4}
\end{figure}

Consequently, we have used the following procedure of extracting
$\tau_{\rm v}$ from the WL magnetoconductance. Firstly, we
analyzed the MR data in sufficiently weak magnetic fields and at
low temperatures. In this regime (the hatched regions in
Fig.~\ref{fig1}), the dephasing occurs at a time scale much
greater than $\tau_{\rm v}$, and we can apply
Eq.~(\ref{HikamiFormula}) for extracting $\tau_{\varphi}$; the
second adjustable parameter, prefactor $\alpha$, appears to be
close to 1. At the next stage, we substitute $\tau_\varphi$ into
the ``ballistic'' formulae
[Eqs.~(\ref{sigma_gen}),~(\ref{sigma_a}),~(\ref{sigma_b})] and
calculate the MC curves in a wide range of fields ($b<1$) for
various $\tau_{\rm v}$. Figure~\ref{fig4} illustrates this
procedure using as an example the same MC data as in
Fig.~\ref{fig2}.   We calculate $\Delta\sigma (b)$ using the
summation technique similar to that described in
Ref.~\cite{mcphail} for single-valley systems.

Figure \ref{fig4} shows that the experimental MC (circles) is
smaller than the MC for a system with two unmixed degenerate
valleys (curve {\it 1}) and larger than MC for a single valley
system (curve {\it 5}). This observation {again} indicates that
the MC in the studied Si MOS structures is affected by valley
mixing. Curves {\it 2}, {\it 3}, and {\it 4} in Fig.~\ref{fig4}
correspond to ${\tau_{\rm v}}/{\tau}$ = 15, 12, and 9,
respectively. Note that in the magnetic field range $b<0.15$,
these three curves, the experimental data, the HLN formula, and
the ballistic result for a single valley system are almost
indistinguishable from each other (see the inset to
Fig.~\ref{fig4}). Therefore, $\tau_{\rm v}$ cannot be reliably
found from the MC in low fields.

Figure~\ref{fig4} shows that the discrepancy between the curve
{\it 5} for a single-valley system and the curves {\it 2,3,4} for
two mixed valleys ($\tau_{\rm v}/\tau=$15, 12, and 9) grows as $b$
increases. This observation has a transparent physical
explanation: with increasing $b$, the typical size of electron
trajectories which contribute to the WL correction diminishes, and
the valley mixing over the time of travel along these trajectories
becomes small when $b>\Phi_0/D\tau_{\rm v}$. As a result, the WL
magnetoconductance in strong magnetic fields approaches the
theoretical prediction for a two-valley system with no intervalley
scattering.

\vspace{0.15in}

We also note that all calculated curves deviate from the
experimental data. As Fig.~\ref{fig4} shows, curve {\it 4}
calculated for $\tau_{\rm v}/\tau=9$ at $b>0.4$ is approximately
parallel to but lower than the experimental data in magnetic
fields $b>0.4$. On the other hand, curve {\it 2} calculated for
$\tau_{\rm v}/\tau=15$ almost coincides with the data in low
magnetic fields $b<0.4$, though deviates substantially from them
in higher fields. The minimal mean-square deviation of the
calculated curve from the data is realized for  $\tau_{\rm
v}/\tau$ =12 (curve {\it 3}).

\begin{figure}
\centerline{ \psfig {figure=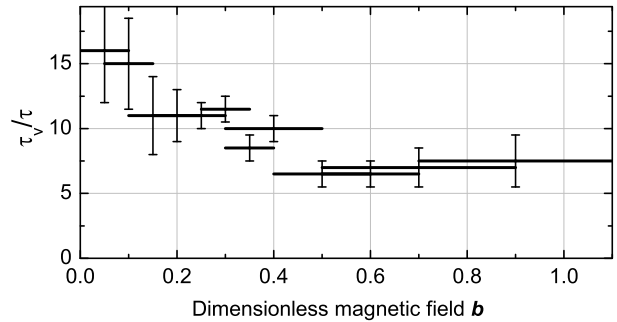,width=270pt,height=160pt}}
\vspace{0.2in}
\caption {Intervalley scattering time determined from fitting the
difference, $\Delta \sigma(b_1)-\Delta \sigma(b_2)$, for the same
magnetoresistance curve as in Fig.~\protect{\ref{fig4}}. The
fitting ranges ($b_1 , b_2$) are shown by the horizontal bars.
$\tau_\varphi/\tau=133$.}
\label{fig5}
\end{figure}

Thus, the value of $\tau_{\rm v}$ depends on the magnetic field
interval ($b_1$,$b_2$) within which the MC data is fitted. {The
($\tau_{\rm v}/\tau$) values, obtained from fitting the difference
$\Delta\sigma (b_1)-\Delta\sigma (b_2)$ as a function of
($b_1$,$b_2$), decrease as $b=(b_1+b_2)/2$ increases (see
Fig.~\ref{fig5})}. This monotonic dependence has been reproduced
for all samples and temperatures. We believe that this apparent
$\tau_{\rm v}(b)$ dependence is an artifact of the fitting
procedure. In all above calculations we assumed $\tau_\varphi$ to
be field-independent. However, $\tau_\varphi$ should depend on a
perpendicular magnetic field \cite{znatauphi}. To the best of our
knowledge, there are neither experimental nor theoretical
systematic studies of this dependence beyond the diffusive limit.
Ignoring this dependence in our fitting could lead to the observed
monotonic variation in $\tau_{\rm v}$ with $b$.

A question arises therefore what range of magnetic fields should
be chosen for $\tau_{\rm v}$ extraction? To answer this question
we have analyzed errors of our method; the resulting
root-mean-square sum of all errors is shown by the error bars in
Fig.~\ref{fig5}. The error analysis is presented in detail in the
Appendix (Sec. \ref{appenA}), where it is shown that neither small
fields ($b<0.1$) nor large fields ($b\sim 1$) should be used for
$\tau_{\rm v}$ extraction. In weak fields, the MC is insensitive
to $\tau_{\rm v}$, whereas in strong fields one approaches the
limits of applicability of the theory developed in
Sec.~\ref{theory}.

Therefore, we conclude that an intermediate range of magnetic
fields is most suitable for extracting $\tau_{\rm v}$. For the
further analysis, we choose the range $b=0.2$ -- 0.4. We have
verified that our conclusions on the temperature and density
dependences of $\tau_{\rm v}$ are not affected if this range is
changed.

\section{Results and discussion}
\label{results}
\subsection{Phase breaking time}

\begin{figure}
 \centerline{\psfig{figure=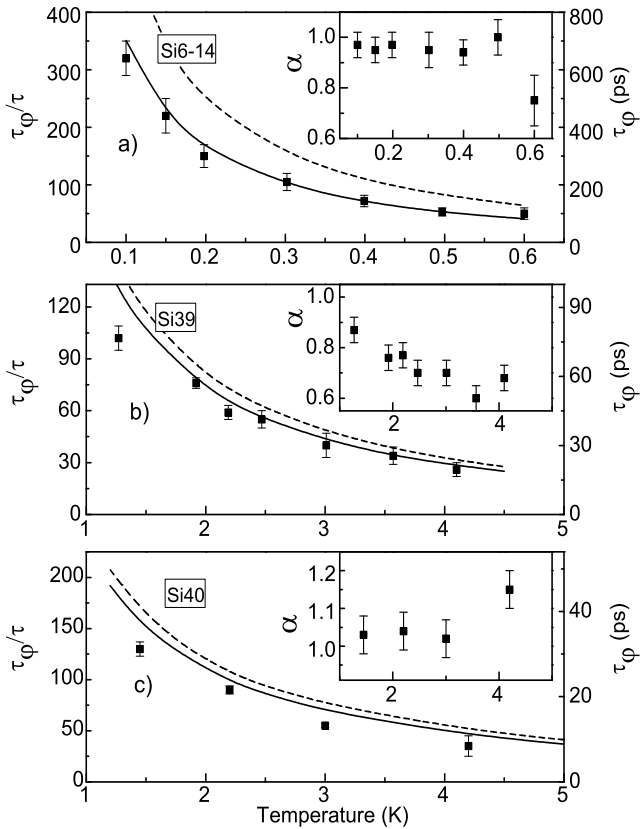,width=270pt,height=360pt}}
 \caption {Temperature dependence of the
extracted $\tau_\varphi$ value in units of $\tau$ (left axes) and
in picoseconds (right axes): a) Si6-14,
$n=9.98\times$10$^{11}$cm$^{-2}$, b) Si39,
$n=29.4\times$10$^{11}$cm$^{-2}$, c) Si40,
$n=33.4\times$10$^{11}$cm$^{-2}$. Solid lines show the
$\tau_\varphi (T)$ dependence predicted by the theory of
interaction corrections \protect\cite{znatauphi} with 15 triplet
terms, dashed line - with 3 triplet terms. The insets show the
corresponding temperature dependences of the prefactor $\alpha$. }
\label{fig6}
\end{figure}

As we have already mentioned, at the first stage of the analysis
we estimated the phase breaking time $\tau_\varphi$. Comparison of
the $\tau_\varphi(T)$ dependences with the theory of interaction
effects \cite{znatauphi} is shown in Figs.~\ref{fig6} a-c. The
uncertainty in the values of $\alpha$ and $\tau_\varphi$ (shown as
the error bars in Fig.~\ref{fig6}) reflects mainly the uncertainty
in $\sigma(b)$ in the weak fields $b <0.01$. The magnitude of the
phase breaking time and its temperature dependence are in good
agreement with the theory for all samples within the studied
ranges of electron density [Si6-14: $n=(0.28-1.5)\times 10^{12}$
cm$^{-2}$, Si39: $n=(2-2.5)\times10^{12}$ cm$^{-2}$, Si40:
$n=(3-4)\times 10^{12}$ cm$^{-2}$]. Note that no adjustable
parameters are involved in this comparison, the Fermi-liquid
parameter $F_0^\sigma$ and the effective mass $m^*$ were obtained
in independent measurements \cite{pudalovflconstants}.

The theoretical curves (solid lines in Fig.~\ref{fig6}) are
calculated following Ref.~\cite{znatauphi} for 15 triplet channels
\cite{finkel'steinPRL}, which implies small valley splitting
($k_{B}T>\Delta_{\rm v}$) and relatively weak intervalley
scattering ($\hbar/\tau_{\rm v} <k_{B}T$). As found from the
analysis of the low-temperature transport and magnetotransport
data \cite{refdeltav}, the condition $k_{B}T>\Delta_{\rm v}$ was
satisfied for samples Si6-14 and Si39 over the major part of the
studied temperature range. Whether or not this condition is
fulfilled for low mobility samples Si39 and Si40 is actually not
important, because the measurements were performed at such high
densities that the amplitude of the triplet term in the
interaction corrections to $\tau_\varphi$ was small in comparison
with the singlet term: changing the number of triplet terms from
15 (two-valley case) to 3 (single-valley case) caused variation of
$\tau_\varphi$ by less than 5\% (dashed line in Fig. \ref{fig6}).
The condition $\hbar/\tau_{\rm v} <k_{B}T$ is violated at
temperatures lower than 0.3 K for sample Si6-14. Still, the
$\tau_\varphi(T)$ data agree better with theory when all 15 rather
than 3 triplet terms are taken into account. {The observed
quantitative agreement of the experimental values of
$\tau_\varphi$ with the theory suggests that $\tau_\varphi$ is
weakly affected by intervalley scattering near the crossover
$k_BT\tau_{\rm v}/\hbar \sim1$}.

\subsection{Prefactor $\alpha$}
\label{prefactorsection} By fitting the weak-field MC data with
the HLN theory, we obtained the prefactor $\alpha$ {that is close
to 1 for all samples (see the insets to Fig.~\ref{fig6}); this
suggests} that the valleys are intermixed on the $\tau_\varphi$
time scale. The decrease of $\alpha$  from 0.9 to 0.6 with
increasing temperature, {obtained for sample Si39   (see the inset
to Fig.~6\,b), we believe, is an artifact,  because} relatively
small values $\tau_\varphi/\tau\sim30$, observed for this sample
at high temperatures, make Eq.~(1) inadequate. The complete theory
described in Sec. \ref{theory} explains that small value of
$\alpha$ is a consequence of a fast phase relaxation. For the same
reason, there is a larger scattering in the values of $\alpha(T)$
for samples Si6-14 and Si40 at the highest temperatures
(Fig.~\ref{fig6}), where $\tau_\varphi$ is small.

It is worth noting that the smallness of prefactor $\alpha$ has
been attributed to the intervalley scattering in
Ref.~\protect\cite{kawaji}. However, the MC data in this
experiment were fitted with the theory~\cite{Kawabata84}, which
does not take into account the non-backscattering correction
Eq.~(\ref{sigma_b}). Our estimates show that for the parameters of
samples studied in Ref.~\cite{kawaji} ($\tau_\varphi/\tau=20$,
$\tau_{\rm v}/\tau=4$, and $b=0.5$--$7$), the non-backscattering
correction contributes  about 50\% to the extracted value of
$\tau_{\rm v}$.

\subsection{Intervalley scattering time: independence of temperature}
\begin{figure}
\vspace{-0.15in}
 \centerline{\psfig{figure=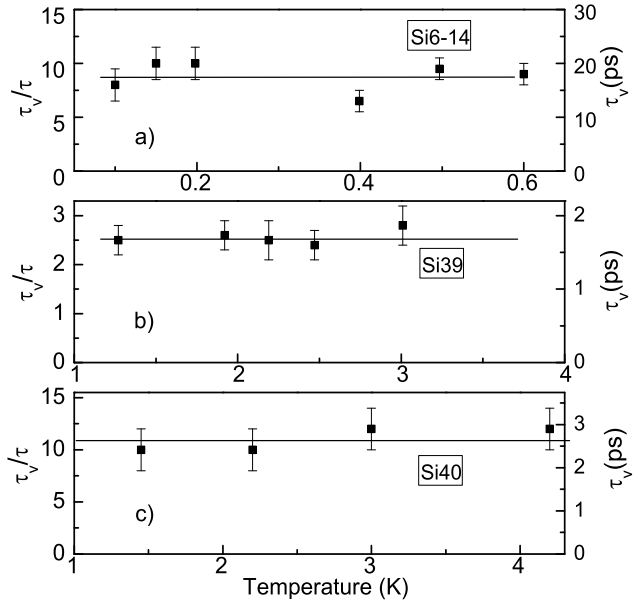,width=3.7in,height=3.7in}}
\caption {Temperature dependence of $\tau_{\rm v}$ in units of
$\tau$ (left axes) and in picoseconds (right axes): a) Si6-14
n=9.98$\times$10$^{11}$cm$^{-2}$, b) Si39
n=29.4$\times$10$^{11}$cm$^{-2}$, c) Si40
n=33.4$\times$10$^{11}$cm$^{-2}$. Solid horizontal lines show the
average $\tau_{\rm v}$. } \label{fig7}
\end{figure}

Following the procedure described in Sec.
\ref{ballisticprocessing}, we have extracted  $\tau_{\rm v}$
 by fitting the WL MC data with  ``ballistic''
Eqs.~(\ref{sigma_gen}),(\ref{sigma_a}),(\ref{sigma_b}).
Figure~\ref{fig7} shows that the values of $\tau_{\rm v}$ are
temperature-independent within the accuracy of our measurements.
{\it This observation suggests that the intervalley scattering is
elastic, i.e. governed by static disorder}. Similar conclusion can
 be also drawn from the fact that $\alpha$ remains close to 1,
while the extracted $\tau_\varphi$ much exceeds $\tau_{\rm v}$ and
grows without saturation as $T$ decreases (see Fig.~\ref{fig6}).
Indeed, were the intervalley scattering inelastic, one would have
observed a prefactor $\alpha\sim2$ because the dephasing would
occur in two valleys independently and the intervalley scattering
would be just an additional dephasing mechanism. In the latter
case, a cut-off of the dephasing time at the level
$\tau_\varphi=\tau_{\rm v}$ is also expected. The two
observations, the absence of the cut-off and $\alpha\approx1$
support the self-consistency of our analysis.

{The intervalley transitions} are not expected  to be inelastic
for the following reason. The intervalley scattering requires a
large momentum transfer comparable to the vector of reciprocal
lattice $2\pi/a\sim 10^8$cm$^{-1}$ ($a$ is the interatomic
distance). At liquid helium temperatures only static disorder can
cause these transitions, as the momenta of electrons
$k_{F}\sim10^{6}$cm$^{-1}$ for the studied range of densities and
phonons $k_{ph}\sim k_BT/(\hbar v_s)$ (here $v_s$ is the sound
velocity) are much smaller than $2\pi/a$. Static disorder can lead
only to elastic scattering because it changes momentum of
scattered electrons but does not change their energy.

\subsection{Intervalley scattering time: density and sample dependence}
\begin{figure}
\centerline{\psfig{figure=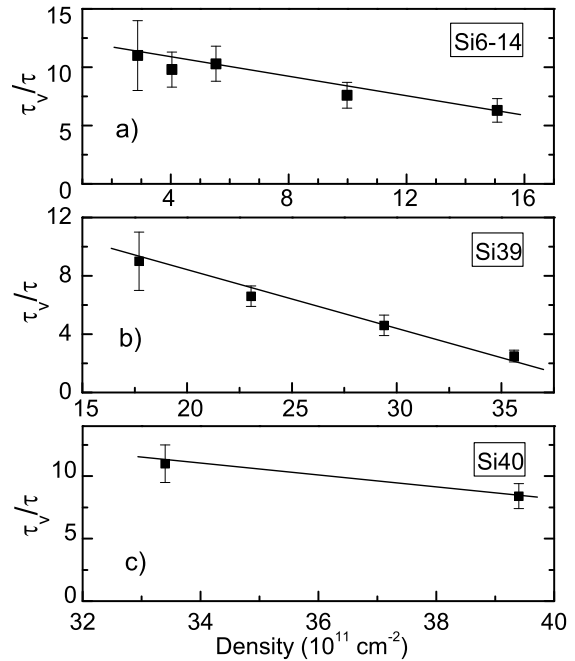,width=240pt}}
\caption {Density dependence of the intervalley
scattering time (averaged over temperature)  for samples Si6-14
(a), Si39 (b) and Si40 (c).} \label{fig8}
\end{figure}

Figure~\ref{fig8} shows the density dependence of $\tau_{\rm v}$
values averaged over the temperature. For all three samples, the
{\it relative} rate of the intervalley transitions (with respect
to the momentum relaxation rate) increases with density. This
points to the {\it dominant role of the Si-SiO$_2$ interface in
the intervalley transitions}. The electron wave function $\Psi$ in
Si MOS structure is positioned mostly in the bulk silicon and
exponentially decays in SiO$_2$ \cite{ando}. When the gate voltage
(and, hence, the density $n$) is increased, the electrons are
``pushed'' towards the Si-SiO$_2$ interface, and the amplitude of
the wavefunction at the interface, $\Psi_0$, increases. The
probability of the interface scattering is proportional to
$|\Psi_0|^2$  and increases with $n$ \cite{ando}; this is in line
with the behavior shown in Fig. \ref{fig8}.

In the experiments we used samples with the mobilities which vary
over a decade. We find no correlation between $\tau_{\rm v}$ and
the mobility for different samples. This suggests that the
intervalley scattering is determined by a sample-specific
interface disorder, namely the surface roughness at the atomic
length scale, which might be different for the samples fabricated
on different wafers. In contrast to the intervalley scattering,
the mobility is governed mostly by impurities in the bulk and by
the interface roughness at a large length scale, $\sim 2\pi/k_{\rm
F}$.

The measured values of $\tau_{\rm v}$  for all samples are within
the interval (3-12)$\tau$, which indicates that the valley index
remains a good quantum number at the time scale $\sim\tau$.

\section{Summary}
\label{summary} To summarize, we have studied the weak
localization magnetoconductance in Si MOS structures over wide
ranges of the electron densities, mobilities, and temperatures. In
order to quantitatively analyze the experimental data, we have
developed the theory of weak localization for two-dimensional {
multivalley systems}, which is valid in both the diffusion and
ballistic regimes. The {theory, which explicitly takes the
intervalley scattering into account, allowed} us to conduct the
first detailed study of the intervalley scattering in the Si MOS
structures. It was found that:
\begin{enumerate}
\item Intervalley scattering in Si MOS structures is an elastic
and temperature-independent process.

\item The ratio $\tau_{\rm v}/\tau$  monotonically increases as
the electron density decreases. This observation suggests that the
intervalley scattering is governed by the disorder at the
Si-SiO$_2$ interface.

\item There is no simple correlation between the intervalley
scattering rate and the sample mobility (or the momentum
relaxation rate); this points to a sample-specific rather than
universal mechanism of the intervalley scattering.

\item The smallness of the prefactor $\alpha\sim1$, that is
obtained from fitting the experimental WL data with the HLN
formula, is a consequence of a fast intervalley relaxation rate
which exceeds the phase relaxation rate.

\item The temperature dependence of the phase relaxation time in
Si MOS structures is in quantitative agreement with the theory of
electron-electron interaction effects in disordered
two-dimensional systems \cite{znatauphi}.

\end{enumerate}

We note that the approach similar to that developed in this paper
can be used for studies of intervalley relaxation in other
multi-valley two-dimensional electron systems, such as AlAs-AlGaAs
heterostructures \cite{shayeganstructures}, Si MOX structures
\cite{simoxstructures}, and Si-SiGe quantum wells
\cite{sigeqwstructures}.

\section{Acknowledgements}
The authors are thankful to I.~V.~Gornyi and G.~M.~Minkov for
illuminating discussions. The research at Lebedev Institute and
Ioffe Institute was supported by RFBR, INTAS, Programs of the RAS,
Russian Ministry for Education and Science, Program ``Leading
scientific schools'' (grants 5596.2006.2, 2693.2006.2, and the
State contact 02.445.11.7346) and Russian Science Support
Foundation. NK and MG acknowledge the NSF support under grant
ECE-0608842. AK acknowledges Education and Research Center at
Lebedev Physics Institute for partial support.

\section {Appendix: Analysis of possible errors in
$\tau_{\rm v}$} \label{appenA} We present here an analysis of
errors in the fitting procedure which determine the size of error
bars in Fig.~\ref{fig5}. As discussed in
Section~\ref{ballisticprocessing}, $\tau_{\rm v}$ was found from
the following equation:

\begin{equation}
\Delta\sigma(b_1)_{\rm EXP}-\Delta\sigma(b_2)_{\rm EXP} =
\label{maineq}
\end{equation}
\vspace{-0.3in}
\[
=\Delta\sigma(b_1,\tau_{\varphi}/\tau,\tau_{\rm_{\rm
v}}/\tau)_{\rm
TH}-\Delta\sigma(b_2,\tau_{\varphi}/\tau,\tau_{\rm_{\rm
v}}/\tau)_{\rm TH}
\]
Here subscript EXP denotes experimental data, subscript TH denotes
calculation using
Eqs.~(\ref{sigma_gen}),(\ref{sigma_a}),(\ref{sigma_b}).
Consequently, the uncertainty in $\tau_{\rm v}$ is determined by
(i) uncertainty
 in $b$, (ii) uncertainty in $\tau_\varphi$ and (iii)
uncertainty in the conductivity. To estimate each contribution to
the error, we varied the corresponding parameter ($b$,
$\tau_\varphi$ or $\Delta\sigma$) within its uncertainty and
determined the variation in $\tau_{\rm v}$ by solving
Eq.~(\ref{maineq}).

The uncertainty $\delta b$ in $b=2B\pi l^2/\Phi_0$ value is
determined by the uncertainty in the mean free path $l$
\cite{ldeterm}. The latter is about 2-3 \% due to the
uncertainties in electron density $n$ and Drude conductivity.
However, $\delta b$ affects $\tau_{\rm v}$ rather weakly for the
following reason: MC in the studied magnetic field range behaves
approximately as $\ln(b)$, therefore
$\Delta\sigma(b_1)-\Delta\sigma(b_2)\sim
\ln(b_1/b_2)=\ln(B_1/B_2)$.

The error related to the uncertainty in $\tau_\varphi$ is
essential in low magnetic fields where magnetoconductance is
sensitive to $\tau_\varphi$. Correspondingly, the error bars in
low fields $b<0.15$  in Fig.~\ref{fig5} are determined
predominantly by the uncertainty in $\tau_\varphi$.

Another source of errors is related to the precision of the
absolute value of WL MC {(``calibration error'')}. Indeed, the
accuracy of our measurements of the absolute magnetoresistance
value is $\sim0.5$\%. Higher order corrections, Maki-Thompson and
DOS corrections\cite{minkovgornyi} can modify MC by approximately
2-3\% (as shown in Ref.~\cite{minkovgornyi}, $\delta (\Delta
\sigma)/ \Delta \sigma\approx 2e^2\rho_{\rm D}/2 \pi^2 \hbar
\approx0.025$). In order to estimate this error we artificially
changed our experimental data by 3\% and studied the corresponding
change in $\tau_{\rm v}$. The error appears to grow in small
magnetic field where magnetoconductance is weakly sensitive to
$\tau_{\rm v}$. Therefore, {\em small fields should not be used
for the extraction of $\tau_{\rm v}$}.  In large magnetic fields
($b\sim 1$) MC becomes again weakly sensitive to $\tau_{\rm v}$,
and the latter error grows as $b$ increases, as shown by the error
bars in Fig. \ref{fig5}. The calibration error is minimal in
intermediate magnetic fields, where MC is most sensitive to
$\tau_{\rm v}$.

Our attempts to analyze the WL MC data in strong fields $b>1$
using Eqs.~(\ref{sigma_gen}),(\ref{sigma_a}),(\ref{sigma_b})
resulted in a large uncertainty of the fitting parameter
$\tau_{\rm v}/\tau$ (large scattering of extracted $\tau_{\rm
v}/\tau$ for various electron densities and temperatures). In
large magnetic fields, there are several other error mechanisms
which are difficult to take into account. For example, at $b\sim
1$, $\tau_\varphi$ differs from its small-field value
\cite{znatauphi}. Moreover, in Ref.~\cite{germanenkonew} the MC
for $b>1$ was shown to behave in a non-universal manner: it
strongly depends on details of scattering potential, whereas our
theory assumes an uncorrelated short-range disorder. Some other
mechanisms of magnetoconductance (such as classical memory
effects, interaction corrections, Maki-Thompson corrections etc.)
may also become essential in large fields where the shape of WL MC
curve flattens. Therefore we believe that the intermediate field
range $b=0.2-0.4$ is optimal for the extraction of the intervalley
scattering rate.


\begin{thebibliography}{99}
\bibitem{ldeterm}
The transport mean free path $l$ was determined from the formula
$\sigma_{\rm D}= (e^2/2\pi \hbar) \times k_{\rm F}l \times n_{\rm
v}$, where $k_{\rm F}=\sqrt{2\pi n / n_{\rm v}}$, $n_{\rm v}=2$
for a two-valley system, $\sigma_{\rm D}$ is the Drude
conductivity which for simplicity was  taken equal to
$\sigma(B=0)$;   this simplification is justified for $\sigma\gg
e^2/2\pi \hbar $.
\bibitem{deltav}
We neglected spin-orbit splittng because no signatures of the
antilocalization were seen in our Si-MOS structures down to the
lowest temperatures. Our theory treats the valleys as equivalent,
assuming $\Delta_{\rm v}\tau/\hbar<1$. As for the experimental
data, from independent estimates \cite{refdeltav} we have verified
that the inequality $\Delta_{\rm v}\tau/\hbar<1$ is satisfied for
all studied Si MOS structures. In general, depending on the valley
splitting, the weak localization should be described either by the
theory developed in the present paper ($\Delta_{\rm v}\tau/\hbar
<1$) or the theory~\cite{averkiev} ($\Delta_{\rm v}\tau/\hbar
\gg1$).
\bibitem{shortrange} As follows from our
SdH measurements for all samples, as well as from numerous other
experiments \protect\cite{ando}, the ratio of the large-angle
scattering time $\tau$ to the quantum time (or any-angle
scattering time) $\tau_q$ is close to unity in the studied range
of electron densities. This result implies that the scattering in
the studied samples is short-range.
\end{thebibliography}
\end{document}